\documentclass[letterpaper,twocolumn,10pt]{article}
\usepackage{usenix-2020-09}

\usepackage{tikz}
\usepackage{amsmath}

\usepackage{filecontents}

\usepackage[linesnumbered, ruled]{algorithm2e}

\begin{document}

\date{}

\title{\Large \bf Scaling Bockchain with Adaptivity\\}

\author{
{\rm Yan Huang, Yu Zhou, Tao Zhu, Yuzhuang Xu, Hehe Wang,}\\
{\rm Weihuai Liu, Jingxiu Hu, Pushan Xiao}\\
{\rm \textit{China UnionPay}}
} 

\maketitle

\begin{abstract}
This paper presents Balloon, a scalable blockchain consensus protocol which could 
dynamically adapt its performance to the overall computation power change. Balloon is
based on a parallel chain architecture combined with a greedy heaviest sub-chain selection strategy. It adopts an inovative block sampling approach to assess the change of block generation rate in the network. By introducing view change mechanism, Balllon is able to dynamically adjust the number of parallel sub-chains. Balloon redefines the concept of block subtree weight with view change in consideration, so that a total order of blocks could be obtained safely. To deal with rapidly increasing block generation rate in the blockchain network, participants of previous Nakamoto-style protocols are required to continuously increase their mining difficulty so as to maintain an expected security gurantee. Balloon, however, could accomadate a fixed difficulty setup and assign superfluous block processing capability to new sub-chains, which makes it more open and also economical.
\end{abstract}

\section{Introduction}
Blockchain has attacted numerous attention since the inception of bitcoin in 2008 \cite{nakamoto2008bitcoin} and 2009 \cite{nakamoto2009bitcoin}. A major drive behind this is the boom of cryptocurrency industry, though it has the potential of optimizing the practice of other industries \cite{novo2018blockchain, sikorski2017blockchain, wang2018understanding} for its efficiency in coordinating between multiple stakeholders. Compared to traditional web service, blockchain could be viewed as a new sercie model under which stakeholders could equally and autonomously participate in the network to achieve certain goals, e.g., bitcoin token transfer or more complex type of transactions defined by smart contract \cite{wood2014ethereum}. At its core, peer-to-peer network and consensus protocol form the backbone of most blockchain systems.

Bitcoin is the first blockchain system that has provided cryptocurrency service to Internet-scale miners. Proof of Work (i.e., PoW \cite{garay2015bitcoin}) is used to reach agreemnt on a total order of blocks. Despite its unsatisfying performance (about 7 transaction per second), it is still supported by miners around the world for its stable scalability and security. However, the safety gurantee is maintained by continuously increasing the mining difficulty over the years (with more partipants joining), which costs too much on the total computation power invested per block. For example, bitcoin consumes approximately 2132.34 kWh per transaction \cite{bitcoinConsumption2022}, about the average consumption of an U.S. household in 70.75 days \cite{powerConsumptionUS2019}. Moreover, higher mining difficulty limits the access of nodes to bitcoin network, making the blockchain more centralized and actually less open. Later blockchains (e.g., ethereum \cite{wood2014ethereum}, conflux \cite{li2020decentralized}, prism \cite{bagaria2019prism}, etc.) also suffer from that.

We present Balloon, a scalable blockchain consensus protocol which could dynamically adapt its block throughput to the overall computation power change. Balloon is constructed on a parallel chain achitecture combined with a variant of GHOST \cite{sompolinsky2015secure} in order to obtain high performance. At its core, Balloon uses an inovative block sampling approach to dynamically evaluates the change of computation power in the network and adjusts the number of its sub-chains (if necessary) with what we call view change mechanism. We reconstruct GHOST safely accross all views so as to achieve final consensus on blocks and transactions. Balloon makes it possible to use a fixed block mining difficulty without impairing the chain security. Its accessibility to nodes with general hardware is promising to create a more open and fair blockchain network.

There are three main obstacles to the design of Balloon. First, it needs an efficient way to assess the overall block generation rate before adjusting its parallel chain. Second, Balloon should not sacrafice its safety during the adjustment to balance superfluous power in the network. Finally,  a total order of blocks across all views should be finally obtained safely. Balloon tackles these obstacles with several new techniques which is described as follows.

\textbf{Dynamic block sampling}. Whenever mining a new block, a miner would search for blocks with a specific clock and time distance from the current one. These sampling blocks would then be added to corresponding block field. Clock is an inner property of block to identify the logical order between different block events. The first sampling block is found along its guider chain (i.e., largest clock direction). After that, other sampling blocks are collected based on its clock and view. Balloon utilizes sampling blocks to estimate block generate concurrency during a specific period (i.e., epoch).

\textbf{View change mechanism}. Balloon periodically detects if there is a potential chain adjustment and responds to significant change of computation power by view change. The process moves Balloon into a new view with different number of sub-chains so as to balance superfluous network power or just enter into a more silent mode with low block throughput. To ensure its safety, Baloon adapts to apparent increasing of computation power in a responsive way and decreasing of it in a conservative way, which could efficiently counter an attacker's manipulation of computation power invested. Moreover, blocks in a new view will have larger clocks than that of its previous views (excluding blocks after anchors) to further strengthen Balloon's safety.

\textbf{Reconstructing GHOST across views}. View change complexifes the total ordering of Balloon since GHOST could not be applied directly and safely. To solve this, Balloon first deserts main sub-chain blocks after anchors (from which view change started) to ensure global consistency acrross views. On the other hand, deserted blocks would still contribute to corresponding block sub-tree weight. Moreover, for those in subsequent views (including both main view and split one), the weight of them would also count for corresponding blocks in a specific view. Based on these observations, we redefine block sub-stree and its weight for Balloon and eventually reach agreement on a total order of blocks.

The remaining of this paper is organized as follows. Section 2 gives an introduction on related work. Section 3 describes some of the key assumptions and goals. Section 4 provides an overview of Balloon and its details are presented in Section 5. Finally, we conclude our work in Section 6.

\section{Related Work}
\textbf{Nakamoto and GHOST}. Nakamoto consensus protocol  is used in Bitcoin \cite{nakamoto2008bitcoin} and many other cryptocurrencies for its scalability in dynamic peer-to-peer (i.e., P2P) networks. It could maintain a steady block processing capability with nodes joining and leaving dynamically. Its agreement is reached on the longest chain of blocks. GHOST \cite{sompolinsky2015secure} is an improvement on the security of Bitcoin, which is partly used in Ethereum. It selects main chain blocks along the direction of heaviest block subtree. Both protocols have probalistic safety gurantee under which honest nodes must occupy more than half of the overall computational power. To keep a stable safety guarantee and performance, block mining difficulty is continuously increased with more nodes joining in the network. Therefore, the cost of computation power on a block will be rapidly increasing, which limits the access of ordinary nodes to the network . It is neither sustainable nor economical in the long run. Despite their unsatisfying performance and sustainbility, they inspired many later works on desiging large-scale consensus protocol in a dynamic P2P network. Balloon is one of them.

\textbf{BFT protocols}. PBFT \cite{castro1999practical} is a typical consensensus protocol that could tolerate byzantine faults \cite{lamport1982byzantine, pease1980reaching}. It is designed as a service-oriented replicated state machine \cite{schneider1990implementing}, which could achieve a deterministic agreement among a known fixed set of participants. For PBFT and its earlier followers like Zyzzyva \cite{kotla2007zyzzyva}, Q/U protocol \cite{abd2005fault} and also its recent variants (e.g., Tendermint \cite{buchman2016tendermint}, HotStuff \cite{yin2019hotstuff}), they scale poorly to large groups (normally with a size of several or dozens of nodes), not to mention open networks. Moreover, they require a deterministic membership known beforehand which is not practical in a dynamical P2P network. Algorand \cite{gilad2017algorand, chen2016algorand} combines a stateless agreement protocol BA* with verifiable random function (i.e., VRF \cite{micali1999verifiable}) to improve its scalability in a open dynamic network. Like traditional BFT protocols, BA* runs by round and waits to collect expected number of votes in every round to achieve a \textit{final} or \textit{tentative} goal. Therefore, the scalability of Algorand is still heavily effected by the size of its committe.

\textbf{Structuring block/transaction tree}. For GHOST and Nakamoto consensus, side chain blocks are deprecated though they occupy a large portion of the network's bandwidth. To better utilize the resource and thus improve the block throughput, protocols of structuring concurrent blocks in the traditional chain were proposed. Inclusive protocol \cite{lewenberg2015inclusive} proposes a general methodology of reconstructing GHOST or Nakamoto chain into a directed acyclic structure (called block DAG) and then generateing a total order of it. Phantom \cite{sompolinsky2021phantom} takes a new ordering approach to Inclusive's block DAG based on the observation that honest nodes in block DAG are well-connected. Conflux \cite{li2020decentralized} is an improvement on Inclusive protocol. It adopts an adaptive weight mechanism to counter liveness attacks on GHOST. Prism \cite{bagaria2019prism} structures blocks into three types (i.e., vote block, proposer block and transaction block) and mines randomly on corresponding types of Nakamoto chains. Voter chains collectively vote for a total order of leader proposer blocks which is then used to order concurrent transaction block events, achieving a high block throughput. Like bitcoin or ethereum, these protocols would gradually increase its mining difficulty and thus cost per block in the computation power with more nodes joining in order to keep a desired safety gurantee. Avalanche \cite{rocket2019scalable} structures the chain into a block-less DAG. It has good scalability for the use of a random sampling approach to reach probalistic agreement on transactions. DAG in Avalanche is used for query batching purpose.

\textbf{Parallel chain architecture}. OHIE \cite{yu2020ohie} is a Nakamoto-style parallel chain protocol with high transaction throughput. Blocks are mined randomly on its symmetric sub-chains. It defines two additional fields (not part of a block), rank and next\_rank, to help blocks grow so it could be totally ordered. Since block rank is determined by its parent block instead of its latest trailing block, the process could not reflect the actual order of block events and its observation of an unbounded trend between sub-chain growth rates makes the case worse. Eunomia \cite{niu2019eunomia} improves OHIE by introducing virtual logical clock to order concurrent blocks. Balloon also utilizes the concept of logical clock except that it includes it within corresponding block field as its inner property. It makes logical clock not "virtual" anymore and simpler to calculate. Moreover, by introducing the concept of guider and guider chain, block clock in Balloon has more natural semantics and usage.

Different from OHIE and Eunomia, a variant of GHOST is used in Balloon as a sub-chain building block to further improve its performance and security instead of using Nakamoto consensus. In practice, both OHIE and Eunomia are not complete because they have not take into account the change of computation power in blockchain. Actually, the security gurantee could not be maintained with continuously increasing nodes participating in the network. A recent work \cite{wang2021securing} on parallel chain mining difficulty adjustment maybe useful for applying OHIE to practical blockchain systems. Besides, they also could not adapt its performance to computation power change like Balloon.

\section{System Model}
Balloon aims to acheive two main goals 1) \textbf{Safety}. With overwhelming probability, honest nodes of Balloon would eventually reach agreement on a total order of blocks and thus their transactions. 2) \textbf{Liveness}. Balloon could continue to make progress by mining and processing new blocks with transactions.

\textbf{Assumptions}. The safety of Balloon does not rely on network synchrony. To ensure a satisfying adaptivity property, however, a weak \textit{d}-synchrony assumption is used. For example, during a enough long period, the network is allowed to be asynchronous for some time discretely, but the total time of  it should be signifcantly smaller compared to that of \textit{d}-synchrony duration. Moreover, during other times, blocks should be received within a known block propagation delay bound \textit{d} after they were sent through gossip protocol. Like many consensus protocols, strong network synchrony is essential for Balloon to gurantee its liveness (i.e., to make progress). Since Balloon is a GHOST-style protocol, we make the same computation power assumption, i.e., adversaries' total block generation rate $q\lambda$ should be less than that of honest nodes $\lambda$, i.e., \textit{q} < 1. Furthermore, Balloon makes an ideal GHOST assumption to increases its adaptivity. That is, there exists an ideal $r_{0}$ = $\lambda_{0} d$ with a fixed difficulty setup, which could ensure GHOST has a satisfying performance and security gurantee. This assumption is practical and $r_{0}$ could also be obtained by experiment on single GHOST chain as well. Actually, adaptivity is aimed to make sure that sub-chains in Balloon could work efficiently and safely like single GHOST, which has been inspected thoroughly and comprehensively over the years. Other assumptions on cryptographic hash functions and sigatures follow common standard, e.g., cryptographic hash functions used in blockchain should be collision-resistant.
\begin{figure*}[htb]
	\begin{center}
		\includegraphics[scale=0.8]{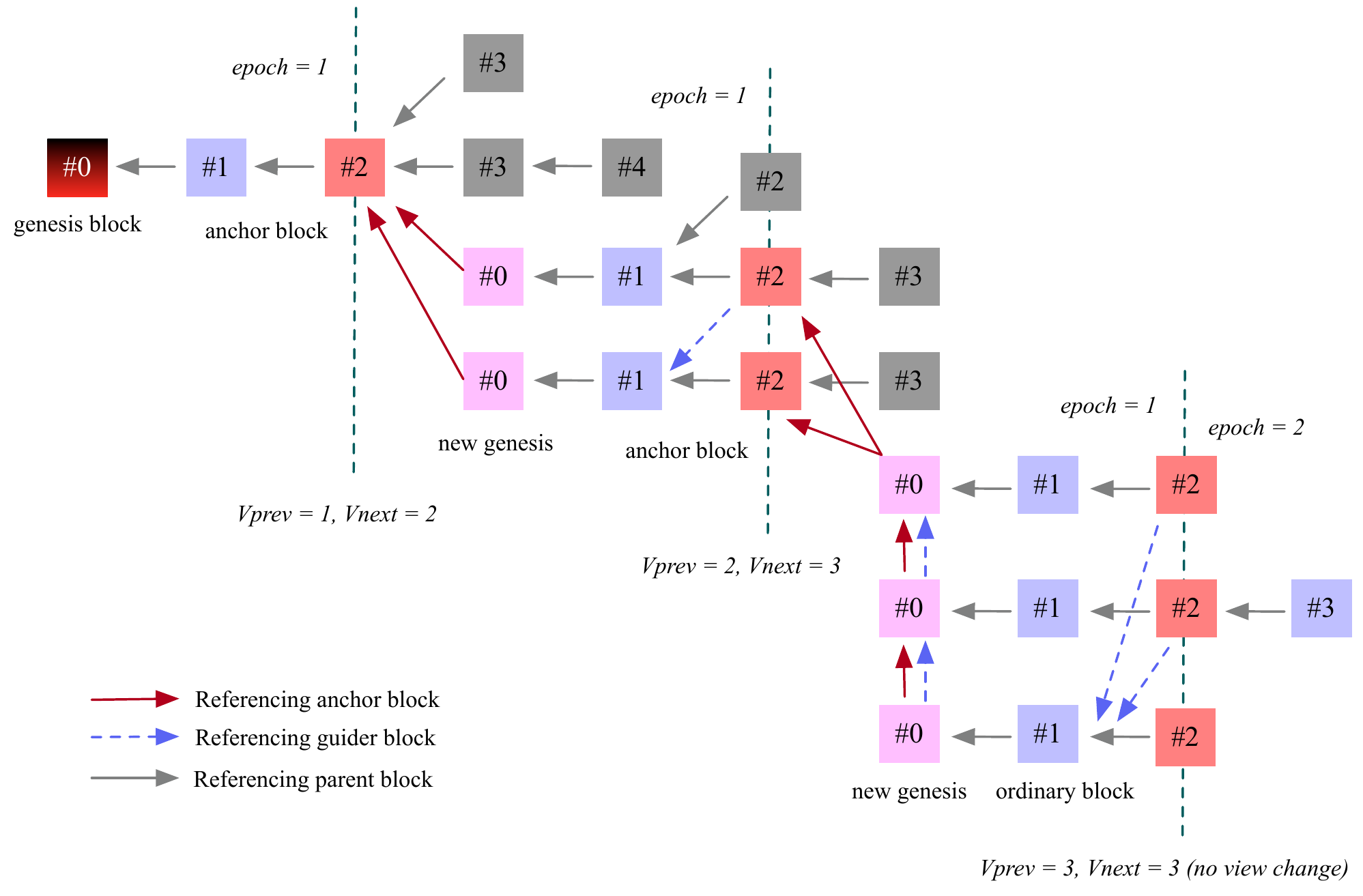}
	\end{center}
\vspace{-0.7cm}
	\caption{\label{fig:architecture} Example of Balloon Chain Structure. It started with a single chain and check if there is a view change every 2 clocks (Note: first epoch is special, see $\S5.3$). $V_{\ast}$ is the main view number. $epoch$ is defined within view and would be reset along with view change. Grey blocks would be deserted in the final total ordering though their weights contribute to Balloon's safety.}
\end{figure*}

\section{Overview}
Balloon inherits the idea of \textit{consensus through competition} from Nakamoto consensus so as to scale blockchain to as many nodes as possible. Every node in Balloon will compete with their
computation power to find a valid block (e.g. whose hash has required number of leading zeros). 
A parallel consensus architecture combined with GHOST is introduced to increase the blockchain throughput. During any stable period (i.e. no chain adjustment happens), all nodes mine blocks randomly and these (if valid) will be assigned to random sub-chains. With more nodes partipating in the blockchain network (or more computation power invested), Balloon will periodically initiate a view change to dynamically increase the number of running sub-chains, or vice versa. Balloon reconstructs GHOST to all blockchain views so as to guarantee overall system safety. We now describe some of the key insights of Balloon.

\textbf{Expanding GHOST to Parallel}. Miners in Balloon generate valid blocks by finding valid PoW puzzles. They start from a single GHOST chain or a multiple of it. We first modify GHOST by calculating the total weights of a block tree instead of the total number of blocks within. The weight of block is defined as its mining difficulty (like in Ethereum). As other parallel blockchains \cite{bagaria2019prism,yu2020ohie}, we add merkle tree root of all latest main sub-chain blocks to the raw block to avoid potential attacks. Later after it is successfully mined, the parent hash and corresponding proof are added accordingly. A common genesis block which records protocol-specific parameters is created before that. 

When mining blocks, a miner will select one block from the current blockchain (only including main chain blocks of those parallel sub-chains) and add its block hash as an inner part of the block. We define this block as the guider block. It is an inborn property of any valid Balloon block. It reflects the \textit{happen-before} relation of blocks \cite{lamport1978time}.  Any guider path of a block dates back to the common blockchain genesis and we call these ordered blocks along the path guider chain. Naturally, we use it to reconstruct Balloon's inner logical clock \cite{lamport1978time}. We thus define block clock as the total number of blocks along the corresponding guider chain (not including the block itself) and add it to corresponding block field as another inner property. 

Before the first view change starts, Balloon orders blocks of multiple sub-chains (corresponding main chain) based on their block clocks. It is simple and natural, since block clock reflects the relative chronological order of block generation just like what block number does in a single chain. For blocks with the same clock, which signifies concurrency of block generation, we could break ties by their hashes or sub-chain ids. When there is a chain adjustment, however, things become complicated and Balloon uses a parallel variant of GHOST to guarantee protocol safety.

\textbf{Sampling Blocks When Mining}. We propose the idea of block sampling to assess overall block generation rate and dynamically adjust the number of parallel sub-chains accordingly. Block Sampling happens during block generation. Whenever mining a new block, a miner would first locate a reference sampling block along its guider chain. In Balloon, we assume there is a known upper bound on block propagation delay most of the time to ensure accuracy of dynamic chain adjustment, though protocol safety doest not rely on it. Obviously, the time difference between the reference sampling block and current one to mine should be larger than the known bound. Practically, we make the difference slightly larger so as to sample blocks more accurately. The miner will then search for all blocks with the same clock as the reference on its sub-chain. Other blocks not in the same view will also be considered (not all of them) if they share the same block clock. They are called sampling blocks of the current block to mine and hashes of them are included into corresponding block field as another inner property. 

We observe that guider chains are symmetric and make progress synchronously with Balloon. Therefore, the guider chain could serve as the timeline for chain adjustment and block clock is the logical time. Balloon could thus be divided into different clock periods in a view along the guider chain and we use epoch to identify them. For any sub-chain in an epoch, we would calculate the average number of sampling blocks per block as an approximation of $\lambda d$. The accuracy of this generation rate depends largely on our knowledge of the propagation delay bound and the epoch length we actually use.

\textbf{Adjusting Chain Democratically}. For PoW-based consensus protocols like Nakamoto or GHOST, they try to avoid double-spend or liveness attack by maintaining a relatively low block generation rate. Most of the time, this is done by continuously increasing the mining difficulty. In Balloon, however, we try to detect the "superfluous" computation power and periodically distribute it onto newly generated sub-chains, or vice versa with the power decreasing.

Balloon triggers chain adjustment based on the number of current sub-chain votes available. A vote is casted if the average number of sampling blocks in an epoch is significantly larger or smaller than the expected one (i.e., $\lambda_{0} d$ in section 3). The significance level is set with a tradeoff between safety and performance. When miners collect more than half of the same sub-chain votes in an epoch, they would prepare a view change. Otherwise, they do nothing and continue to the next epoch. For adversaries, they should occupy more than half of the network computation power to force a view change.

\textbf{Determining Chain Structure of New View}. Before making the adjustment, miners first determines the number of sub-chains in the new view. Two dynamic strategies are available, namely, consertive strategy and responsive one. In the first one, miners always add or delete fixed number of sub-chains after the switch. It is conservative because adversaries could not apparently impair system security by first sharply reducing its overall computation power during one epoch (which forces a chain adjustment) and then restoring it in the next view. On the other hand, the responsive strategy requires the number of sub-chains in the next epoch be totally determined by the significance level of the average numnber of sampling blocks. It could adapt the chain more quickly to sharp change of computation power and allow better performance capacity after an significant increase. Balloon combines two of them by adopting conservative strategy when the number of sub-chains are decreasing and responsive one when there will be an increase.

\textbf{Binding New Genesis to Anchors}. For convenience, we use the concept of view to distinguish between different parallel chain structures, i.e., before and after chain adjustment. We define anchor block as the last main sub-chain block in a epoch after which a view change is prepared. The result of chain adjustment is a complete list of new genesis blocks for the new view. Different from normal block mining, a newly mined genesis block references all anchor blocks in the previous view explicitly or implicitly. It could either include corresponding block hashes to its field directly or references some valid new genesis block indirectly. This way, blockchain of different views are thus connected as a whole. 

After that, Balloon enters into a new view and miners will mine from new genesis blocks just as they did in previous one. The guider of a new genesis is the anchor block with the largest clock. New mining blocks afterwards will only reference those of the same view as their corresponding guiders.

\textbf{Total Order of Blocks and Transactions}. 
The view change process moves Balloon into a new view. Block clocks are higher than that of corresponding prevous view, which makes it still safer to order blocks based on clocks. To guarantee protocol safety, we reconstruct GHOST with observation that the weight of any block in subsequent views should contribute to the total weight of corresponding anchor block tree. One view after another, we search for main sub-chain blocks using the reconstructed GHOST and thus generate an ordered list of them based on clocks. Finally, a total order of blocks across all Balloon views is obtained. For transactions within, we handle them by corresponding block order and just skip conflict transactions, e.g., repeated ones. 

\section{Balloon Consensus Design}
In normal cases, the structure of Balloon is stable and newly mined blocks will be assigned randomly onto sub-chains. During this period, miners continuously sample certain past blocks. They use the estimated sampling block generation rate to decide whether to adjust the chain structure at the end of an epoch. Besides, the new structure is determined based on it if a chain adjustment bill is "passed". Valid new genesis blocks should be first generated before Balloon could finally mining in the new view. For blocks in different views, Balloon further expands the sub-chain GHOST rule to reach agreement on the total order of them without imparing  its safety.
\subsection{Basic Parallel Chain Mining}
A typical mining procedure for normal blocks within a view is presented in Algorithm 1. It starts with the latest main sub-chain blocks $B_{m}$ of current view derived from Algorithm 6 ($\S5.4$). $B_{m}$ serves as snapshot for generating merkle proof of this block later. Honest blocks always determine \textit{b.guider} based on their $B_{m}$ (line 2, e.g. $b_{x} \in G_{p}$ is inappropriate) in case of possible clock attack where an adversary may continuously expanding its guider chain and send them to connected nodes causing transaction delay or even rejection.

\begin{algorithm}
	\caption{Basic Balloon mining process.}
	\KwIn{$G_{p}$: block graph of current parallel chain,  \\\qquad\quad$B_{m}$: list of latest main sub-chain blocks}
	\KwOut{b: a newly mined block}
	\textit{$b.root \leftarrow merkle\_root(B_{m})$}\\
	\textit{$b.guider \leftarrow \mathop{\arg\max}\limits_{b_{x} \in B_{m}} b_{x}.clock$}\\
	\textit{$b.clock \leftarrow b.guider.clock + 1$}\\
	\textit{$b.samples \leftarrow Sample(G_{p}, b)$}\\
	\While{difficulty(b) < Diff$_{required}$}{
		\textit{update(nonce)}
	}
	\textit{b.nonce $\leftarrow$ nonce}\\
	\textit{$h_{c} \leftarrow chain\_hash(b)$}\\
	\textit{$n_{v} \leftarrow \mid\mid B_{m} \mid\mid$}\\
	\textit{$sid \leftarrow h_{c} $ \% $ n_{v}$}\\
	\textit{$(b.parent, b.proof) \leftarrow merkle\_proof(B_{m}, sid)$}\\
	\textit{$b.number \leftarrow b.parent.number + 1$}\\
	~\\
	\textbf{return} \textit{b}
\end{algorithm}

The introduction of block sampling (line 4, detailed in $\S5.2$) is what Balloon mining differentiates most from other consensus protocols. Hashes of corresponding sampling blocks will be included within. $Diff_{required}$ is the target difficulty a new block should satisfy. Different from other protocols relying on computation power, Balloon adopts a fairer strategy by using a fixed difficulty setup ($Diff_{required}$ of line 5) which is relevant with expected hardware processing ability on participating nodes. "Superfluous" computation power crowding in will be periodically balanced by view change process which will be discussed in $\S5.3$. Chain hash $h_{c}$ is calculated by applying hash function to an ordered fields of \textit{b} except its parent and proof field (different from block hash). Sub-chain id \textit{sid} of our newly mined block is derived with a mod operation based on chain hash $h_{c}$ and current number of sub-chain $n_{v}$. Finally, \textit{b}'s parent and its corresponding merkle proof are retrieved to assemble a complete block.
\subsection{Dynamic Block Sampling}
Most Nakamoto-style single chain protocols (e.g. bitcoin, ethereum, etc.) adapt to the network's increasing computation power by adjusting only its main chain growth rate, which could not work correctly in a globally high block generation rate scenario. Balloon introduces block sampling to continuously monitor the variance of overall block generation rate and thus the computation power invested in the whole blockchain network. Algorithm 2 details how sampling blocks are collected.
\begin{algorithm}
	\caption{Block sampling in Balloon.}
	\KwIn{\textit{b}: block to make sampling for, \\ \qquad\quad$G_{p}$: block graph of current parallel chain}
	\KwOut{\textit{S}: sampling blocks found}
	\textit{$b _{h} \leftarrow$ b.guider}\\
	\textit{$GC(b) = GC(b _{h}) \circ b _{h}$ } // Trace until initial genesis\\
	\textit{Anchors(b) = Anchors(Genesis(b))}\\
	\textit{$SameView( b_{1} , b _{2}) = SameAnchors(b _{1}, b _{2})$}\\
	\textit{$ SP(b) = \{b_{s} \mid Diff_{clock}(b_{h}, b_{s}) \geq m $ \&\& $Diff_{time}(b_{h}, b_{s}) \geq n*D, b_{s} \in GC(b)\}$ }\\
	\textit{$b_{st} \leftarrow \mathop{\arg\max}\limits_{b_{x} \in SP(b)} b_{x}.clock$}\\
	\textit{$B_{c} \leftarrow \{b_{c} \mid b_{c}.clock = b_{st}.clock $ \&\& $ SameView(b_{c}, b_{st}) $ \&\& $SameChain(b_{c}, b_{st}), b_{c} \in G_{p}\}$}\\
	\textit{$B_{p} \leftarrow \{b_{p} \mid b_{p}.clock = b_{st}.clock$ \&\& $SameView(b_{p}, b_{st}) \rightarrow false\}$}\\
	\textit{$B_{ps} \leftarrow \{b_{ps} \mid SameChain(b_{ps}, b_{st}), b_{ps} \in B_{p})\}$}\\
	\textit{$ S \leftarrow B_{c} \cup B_{ps} $}\\
	~\\
	\textbf{return} \textit{S}
\end{algorithm}

As shown in Algorithm 2, \textit{GC(b)}, the guider chain of block \textit{b}, is an ordered list of blocks along its guider path dating back to the globally configured genesis $g_{0}$. Obviously, all $GC(\ast)$ begin from $g_{0}$. We define anchor blocks of any block \textit{b} as those of its corresponding genesis along the parent path. A genesis's anchors are blocks in the previous view from which new genesis is mined (see Algorithm 4). We call blocks with the completely identical set of anchors are \textit{in the same view} (line 4). Obviously, blocks having the same view number are not necessarily in the same view.

The goal of block sampling is to capture the average level of block concurrency in the Balloon network. For single chain architecture, this is achieved by calculating the number of blocks of the same block height. In Balloon, however, we measure block concurrency by concurrent blocks per clock. Actually, if we reduce Balloon to single chain mode, block clock and height merge completely with each other. We now start by determining the first expected sampling block $b_{st}$ along its guider chain (i.e. timeline). After that, we will then find its counterparts. 

Under weak d-synchronous network, there are two main factors that directly influence which $b_{st}$ we may use, namely, sampling accuracy and freshness. On one hand, too old blocks could not reflect correctly the recent change of network computation power. On the other hand, fresh blocks are usually not completely received locally. Lines 5 and 6 describe this tradeoff. $Diff_{clock}$ and $Diff_{time}$ are two functions calculating clock and time difference between blocks, respectively. \textit{D} is the upper bound of block propagation delay. The $b_{st}$ we use should at least \textit{m} clocks away from \textit{b}'s guider and \textit{n*D} before it ($m, n \geq 1)$).

$B_{c}$ is a set of blocks sharing the same view, sub-chain id, and clock with $b_{st}$. $SameChain(b_{1}, b_{2})$ gives whether $b_{1}, b_{2}$ have the same sub-chain id using $n_{v}$ of $b_{2}$'s view. Balloon also sample blocks of different views from $b_{st}$ but share the same clock and could also map to $b_{st}$'s current sub-chain (i.e. $B_{ps}$ in line 9). Finally, $B_{c}$ and $B_{ps}$ together constitute \textit{b's} sampling blocks S (i.e., \textit{b.samples}). Balloon utilizes S to approximate the block concurrency level of a single sub-chain.
Besides, it is worthy to note that attacks on the sampling accuracy and freshness of a single block will do little harm to Balloon's safety, since only statistical behavior of block samples is utilized, as discussed in the next subsection. In practice, however, we could set an upper limit on the number of sampling hashes \textit{b.samples} could include, so as to counter attacks of flooding too high sampling payload into a block to mine.

\subsection{Adaptive Chain Adjustment}
Balloon will dynamically adjust the number of sub-chains if it detects significant changes in the network's computation power. We use the concept of view change to distinguish status before and after this adjustment.

\textbf{Triggering chain adjustment}. Algorithm 3 shows how Balloon detects a potential view change. The process starts with genesis blocks of current view and ends with the latest main sub-chain blocks in the same view. It regularly collects samples of all sub-chains during a deterministic period called \textit{epoch}. The length of an epoch is measured by clock (e.g., $C_{p}$ in line 8), which is more responsive than the real time. For the first epoch, the range of clocks could be different (line 17), since the initial clocks of genesis blocks on different sub-chains may vary from each other. The upper bound of the first epoch range is decided by the largest clock of them (i.e., $c_{g}$ in line 2). After the first epoch, different sub-chains share the same epoch range (line 8, 9, 18). The setting of epoch length $C_{p}$ is a tradeoff. Bigger epoch distance (measured by clocks) means more block samples available and better sampling accuracy, but it also brings poor responsiveness and thus bad adaptivity. Considering views may be split from different nodes, it is obvious that the concept of epoch is only meaningful in a specified view scenario as seen in Fig 1.
\begin{algorithm}
	\caption{Triggering chain adjustment.}
	\KwIn{$B_{m}$: list of latest main sub-chain blocks of current view, $B_{g}$: list of genesis blocks of current view corresponding to $B_{m}$}
	\KwOut{Whether to trigger view change}
	\textit{$n_{v} \leftarrow \mid\mid B_{g} \mid\mid$}\\
	\textit{$c_{g} \leftarrow max\{b_{g}.clock \mid b_{g} \in B_{g}\}$}\\
	\textit{$epoch \leftarrow 1$}\\
	\textit{$undone \leftarrow true$}\\
	\While{undone}{
		\textit{$Vote \leftarrow (0, 0, 0)$}\\
		\textit{$(B_{a}, R) \leftarrow (\phi, \phi)$}\\
		\textit{$c_{s} \leftarrow c_{g} + (epoch-1) * C_{p}$} \\
		\textit{$c_{t} \leftarrow c_{s} + C_{p} - 1$}\\
		\textit{$n \leftarrow 1$}\\
		\While{$n \leq n_{v}$}{
			\textit{$b \leftarrow B_{m}(n)$}\\
			\textit{$c_{b} \leftarrow b.clock$}\\
			\uIf{$c_{b} < c_{t}$}{
				$undone \leftarrow false$\\
				\textbf{break}\\
			}
			\lIf{epoch = 1}{$c_{s} \leftarrow B_{g}(n)$}
			\textit{$Ancests(b) = Ancests(b.parent) \cup b$}\\
			\textit{$B_{s} \leftarrow \{b_{s} \mid c_{s} \leq b_{s}.clock \leq c_{t}, b_{s} \in Ancests(b)\}$}\\
			\textit{$b_{a} \leftarrow  \mathop{\arg\max}\limits_{b_{s} \in B_{s}} b_{s}.clock$} \\
			\textit{$B_{a} \leftarrow B_{a} \cup b_{a}$}\\
			\textit{$total \leftarrow \sum_{b_{s} \in B_{s}} \mid\mid b_{s}.samples \mid\mid$}\\
			\textit{$rate \leftarrow  total /  \mid\mid B_{s} \mid\mid $}\\
			\textit{$\alpha \leftarrow \mid\mid rate - r_{0} \mid\mid / r_{0}$}\\
			\eIf{$\alpha \leq \alpha_{0}$}
			{
				\textit{$Vote(0) \leftarrow Vote(0) + 1 $}\\
				\lIf{$Vote(0) > n_{v} / 2$}{\textbf{break}}
				
			}{
				\textit{$R \leftarrow R \circ rate$}\\
				\eIf{$rate > r_{0}$}{
					\textit{$Vote(1) \leftarrow Vote(1) + 1 $}\\
				}{
					\textit{$Vote(2) \leftarrow Vote(2) + 1 $}\\
				}
			}
			\textit{n $\leftarrow$ n + 1}
		}
		\textit{$n_{a} \leftarrow \mid\mid B_{a} \mid\mid$}\\
		\uIf{$ n_{a} = n_{v}$ \&\& $(Vote(1) > n_{v} / 2 \mid\mid Vote(2) > n_{v} / 2)$}{
			\textit{$vote\_up \leftarrow Vote(1) > n_{v} / 2 $}\\
			\textit{$startViewChange(B_{a}, R, vote\_up)$}\\
			\textbf{break}
		}
		\textit{epoch $\leftarrow$ epoch + }1 \\
	}
	\textbf{return}
\end{algorithm}

For every single sub-chain, Balloon collects block samples and make independant vote decesions. $B_{s}$ is the set of blocks to use for current epoch, based on which we define a \textit{potential anchor block} as one with the largest clock in $B_{s}$. It is \textit{potential} because a real anchor should be used for a real view change process but it is yet known as of now. \textit{rate} reflects the block concurrency level over a certain period (e.g. block propagation delay or other time presets in line 5 of Algorithm 2). $r_{0}$ is the corresponding reference level under which a single GHOST sub-chain could achieve satisfying performance and security gurantee. $\alpha_{0}$ is the minimum significance level of $rate$ change that Balloon could tolerate, beyond which a view change vote (for adapting to higher or lower computation power) is casted. It is introduced because frequent view changes would incur more risks of being attacked and meanwhile decrease the overall Balloon chain performance (e.g. switching cost, split view, etc.). $Vote$ records all types of votes casted by sub-chains and remembers them seperately (i.e., $Vote(0)$ for no change, $Vote(1)$ for decreasing $rate$, and $Vote(2)$ for increasing $rate$). 

The final view change decision of an epoch is then made when enough of its sub-chains \textit{cast} their votes. If more than half of sub-chain votes have been casted to support keeping the status quo (i.e. no change), there would no view change at the end of this epoch (line 27). On the contrary, when more than half of them unanimously vote for decreasing rate or the other, a view change process will be initiated but under one extra condition that all sub-chains have cast their ballot (line 39). This extra requirement is essential otherwise miners may select anchors randomly and thus enter into different views, causing blockchain inconsistency. Finally, Balloon begins its view change with consistent anchors $B_{a}$ and corresponding sampling rates $R$.

\textbf{Entering into new view}. A fresh new view setup consist of two components, the number of sub-chains $n_{v+1}$ in the new view and corresponding genesis blocks to beigin with (e.g., $b_{g}$), as shown in Algorithm 4. To determine $n_{v+1}$, Balloon adopts two different strategies with an emphasis on responsiveness and security, respectively. On one hand, if the overall computation power is significantly increasing (i.e., $vote\_up$ is true), $n_{v}$ is set to quickly adapt to this change. \textit{max(R)} corresponds to sub-chain with the largest change and Balloon directly follows it. An alternative approach includes adapting only to the average computation power change. It is suboptimal becasue the security of sub-chains beyond this level is compromised. On the other hand, a conservative strategy is used to cope with apparent decrease in the overall power change. $\alpha_{1}$ is the maximum $rate$ change level allowed during a view change. It is preset and determines the lower bound on $n_{v+1}$ so as to resist attacks by malicious nodes like reducing its computation power in a short time. $r_{m}$ further restrains how far $n_{v+1}$ could deviate from $n_{v}$. In both scenarios, $n_{v+1}$ is decided by $n_{v}$ and specified rate change level (line 3, 6). 
\begin{algorithm}
	\caption{View change process.}
	\KwIn{$B_{a}$: anchor blocks, R: rates beyond allowed level $\alpha_{0}$, $n_{v}$: number of sub-chains of old view, \textit{vote\_up}: true if vote for rate high.}
	\KwOut{$b_{g}$, genesis block of new view}
	$n_{v} \leftarrow \mid\mid B_{a} \mid\mid$\\
	\eIf{vote\_up}{
		\textit{$n_{v+1} \leftarrow \lceil n_{v} * \dfrac{max(R)}{r_{0}} \rceil$}\\
	}{
		\textit{$r_{m} \leftarrow max\{r \mid r < r_{0}, r \in R\}$}\\
		\textit{$n_{v+1} \leftarrow \lceil n_{v} * max\{\dfrac{r_{m}}{r_{0}}, 1- \alpha_{1}\} \rceil$}\\
	}
	\textit{$b_{g}.root \leftarrow null$}\\
	\textit{$(b_{g}.parent, b_{g}.proof) \leftarrow (null, null)$}\\
	\textit{$B_{g} \leftarrow \{b \mid b.anchors \rightarrow B_{a} \mid\mid (b.parent \rightarrow null $ \&\& $b.guider \in B_{g})\}$}\\
	\eIf{$B_{g} \rightarrow \phi $}{
		\textit{$b_{g}.anchors\leftarrow B_{a}$}\\
		\textit{$b_{g}.guider \leftarrow  \mathop{\arg\max}\limits_{b_{a} \in B_{a}} b_{a}.clock$} \\
	}{
		\textit{$b_{g}.anchors\leftarrow \phi$}\\
		\textit{$b_{g}.guider \leftarrow \mathop{\arg\max}\limits_{b \in B_{g}} b.clock$} // better liveness\\
	}
	\textit{$b_{g}.clock \leftarrow b_{g}.guider.clock + 1$}\\
	\textit{$b_{g}.samples \leftarrow Sample(G_{p}, b_{g})$}\\
	\While{difficulty($b_{g}$) < required\_diff}{
		\textit{update(nonce)}
	}
	\textit{$b_{g}.nonce \leftarrow nonce$}\\
	\textbf{return} \textit{$b_{g}$}
\end{algorithm}

Genesis blocks are generated in a way similar to normal block mining except some new characteristics. First, a genesis block should include anchors $B_{m}$ as one of its inner property (i.e., \textit{b.anchors}) and the one with largest clock as its guider if it is the first one of its view. Second, for later genesis block to mine, it would choose existing ones with the largest clock as its guider and anchors is then defined as that of the guider's. Third, snapshot root, parent, and its proof are meaningless fields for new genesis. Actually, they could find all of them in their anchors. Finally, new genesis blocks continue to do the sampling job. It seems a little trickly for cross view sampling, but the rules and rationale presented in $\S5.2$ still apply. After all sub-chains have its corresponding genesis, honest nodes would stop genesis block generation and start to mine normal blocks. 

\subsection{Global Consensus on Blocks}
Compared to GHOST protocol, the total ordering of blocks in Balloon is mainly compexified by mechanisms of block sampling and chain adjustment, also mixed with the parallel chain achitecture. Balloon first simplifies the process by redefining the subtree weight of blocks on a global scale in terms of possible view change and split. After that, it deals with blocks one epoch after another.
\begin{algorithm}
	\caption{Block subtree weight redefinition.}
	\textit{Children(b) =}\{ \textit {b$ _{c} $ $\mid$ b$ _{c} $.parent $\rightarrow$ b}, $b_{c} \in G_{p}$\} \\
	\textit{Offspr(b) = $\bigcup$ $_{b_{c} \in Children(b)}$ Offspr(b$_{c}$) $\cup$} \{b\}\\
	\textit{Anchors(b) = Anchors(Genesis(b))}\\
	\textit{Reformers(b) =} \{ \textit {b$ _{a} $ $\mid$ isAnchor(b$ _{a} $) $\rightarrow$ True, b$ _{a}$ $\in$ Offspr(b)}\} // Mabye multiple anchors per route\\
	\textit{Changers(b) =} \{ \textit {b$ _{g} $ $\mid$ isGenesis(b$ _{g} $) $\rightarrow$ True \&\& Reformers(b) $\cap$ Anchors(b$ _{g} $) $\neq$ $\phi$ }, $b_{g} \in G_{p}$\} \\
	\textit{Successors(b) =}  \textit{$\bigcup _{b_{r} \in Changers(b)} Offspr(b_{r}) \cup Successors(b_{r})$}\\
	\textit{Supporters(b) = } \{ \textit {b$ _{s} $ $\mid$ SameChain(b$ _{s} $, b), b$ _{s}$ $\in$ Successors(b)}\}\\
	\textit{SubTB(b) =} $Offspr(b_{r})$  $\cup$ \textit{Supporters(b)} \\
	~\\
	\textit{SubTW(b) = $\sum Weight(b_{w}), b_{w} \in SubTB(b)$ }  \\
\end{algorithm}

\textbf{Redefining block subtree}. Our key insight is that blocks in subsequent views should also contribute to the stability and security of blocks in the previous view. Steps of how we retrieve Balloon block subtree weight \textit{SubTW(b)} are given in Algorithm 5. \textit{Offspr(b)} is the set of all blocks that could trace back to \textit{b} along its parent path. It is obvious that blocks in \textit{Offspr(b)} are in the same view since genesis blocks do not have \textit{parent} field set (Algorithm 4). Recall that we have defined anchors of a genesis block as blocks within its $anchors$ field or that of its guider. Based on that, we could define anchors of any block (i.e., \textit{Anchors(b)} in line 3) as that of its corresponding sub-chain genesis. $Reformers(b)$ is a set of anchor blocks in $Offspr(b)$. A block is an anchor if it is a potential anchor (line 20, Algorithm 3) and there is at least one valid view change based on it (Algorithm 4). $Changers(b)$ is a set of genesis blocks who has anchors within $Reformers(b)$. $Successors(b)$ is a set of all blocks in the following views which may contribute to \textit{b}'s weight. It is defined recursively as all blocks that may trace back to \textit{b} through view change relation. After that, from the view of \textit{b}, we choose those that could map to \textit{b}'s current sub-chain which we call \textit{Supporters(b)}. As explained in Algorithm 2, $SameChain(b_{1}, b_{2})$ checks whether $b_{1}$ and $b_{2}$ have the same sub-chain id based on the total number of sub-chains in \textit{b}'s view. Finally, we could define subtree blocks of \textit{b} (i.e., \textit{SubTB(b)}) as its \textit{Offspr} plus \textit{Supporters}. Therefore, the block subtree weight (i.e., \textit{SubTW(b)}) is the accumulated weight of blocks within that. It is noted that, just like its counterpart in GHOST, $SubTB(b)$ is directly derived from \textit{b}'s position in the valid chain graph and does not rely on any undeterminitic state info (e.g., current view). Therefore, it is feasible to be used in Balloon chain ordering.
\begin{algorithm}
	\caption{Total order of blocks.}
	\KwIn{Block graph of current parallel chain $G_{p}$}
	\KwOut{An ordered list of blocks $C$}
	\textit{$v \leftarrow 1$} \\
	\textit{$N_{v} \leftarrow n_{0}$} \\
	\textit{$G_{v}(\cdot) \leftarrow g_{0}$} \\
	\textit{C $\leftarrow$ $\phi$} \\
	\While{True}{
		\textit{$vc \leftarrow false$} \\
		\textit{$C_{v} \leftarrow \phi$} \\
		\textit{A($\ast$) $\leftarrow$ G$ _{v} $($\ast$)} \\
		\While{True}{
			\textit{n $\leftarrow$ 1} \\
			\While{n $\leq$ N$ _{v} $}{
				// Last potential anchor or genesis\\
				\textit{b $\leftarrow$ A(n)}\\
				\textit{abk $\leftarrow$ null} // Next potential anchor \\
				\While{b $\neq$ null}{
					\textit{C$ _{v} $ $\leftarrow$ C$ _{v} $ $\circ$ b} \\
					//null if not found\\
					\textit{$b_{m} \leftarrow \mathop{\arg\max}\limits_{b_{c} \in Children(b)} SubTW(b_{c})$}\\
					\eIf{isPotentialAnchor(b$ _{m} $) $\to$ False}{
						\textit{b $\leftarrow$ b$ _{m} $}\\
					}{
						\textit{abk $\leftarrow$ b$ _{m} $}\\
						\textbf{break}
					}
				}
				// Update potential anchor block (if any)\\
				\textit{A(n) $\leftarrow$ abk}\\
				\textit{n $\leftarrow$ n+1} \\
			}
			// View change check\\
			\eIf{checkVC(A) $\to$ True}{
				\textit{Update(G$ _{v+1} $, N$ _{v+1} $)}\\
				// Add anchors of old view\\
				\textit{$C_{v} \leftarrow C_{v} \cup A(\ast)$}\\
				\textit{vc $\leftarrow$ True}\\
				\textbf{break}\\
			}{
				// All blocks processed, exit\\
				\textbf{if} \textit{A($\ast$) $\to$ null} \textbf{then} \\
				\quad \textbf{break}\\
			}
		}
		\textit{SortBlocks(C$_{v}$)} // By its clock and hash\\
		// Collect blocks or current epoch\\
		\textit{C $\leftarrow$ C $\circ$ C$ _{v} $}\\
		\lIf{$vc \rightarrow false$}{\textbf{break} // task done}
		\textit{v $\leftarrow$ v+1}\\
	}
	\textbf{return} \textit{C}\\	
\end{algorithm}

\textbf{Ordering blocks across views}. Block ordering in Balloon is different from other parallel chain protocols in mainly three aspects. First, Balloon orders blocks by view and only main views (compared to view split) are considered. Since view change process (see Algorithm 4) ensures clocks are increasing as a whole for blocks in the subsequent view, it does not break the protocol's safety. Second, main sub-chain blocks are retrieved by applying a variant of GHOST independantly which utilizes $SubTW$ (see Algorithm 5). Finally, in a specific view, blocks after anchors (i.e., \textit{Offspr(b)} except anchor b) will not be ordered anymore though they still contribute to the overall security (by block weight) of the protocol. After that, a consistent order of blocks could be derivated by sorting them based on their block clocks and hashes. Algorithm 6 details how a total order of blocks is obtained from the ground up given a full picture of current block graph $G_{p}$ and some initial setups. 

As shown in Algorithm 6, $v$ is the current view number to differentiate from other main views, $N_{v}$ is the number of sub-chains in main view $v$, $G_{v}$ is a set of corresponding genesis blocks which is initially set to be the same (i.e., $g_{0}$), and $C$ is the total order of blocks we would like to get. As mentioned earlier, Balloon handles blocks in a view order and will generate a view-wise order of blocks $C_{v}$ whose contactention constitutes the expected $C$. In a specific view, however, blocks are processed one epoch after another. Therefore, at the start of an epoch, the starting blocks of corresponding sub-chains are either geneses (first view, line 8) or potential anchors (line 19, 26). For a specific sub-chain in an epoch, selection of main chain blocks (line 18) follows greedy heaviest strategy with the use of new block subtree weight defined in Algorithm 5. After every epoch, we check whether there is a view change to be triggered (i.e., $CheckVC$, detailed in Algorithm 3). We would update our view and genesis if necessary (check against Algorithm 4) and then continue to process blocks in the new view until the end. $C_{v}$ contains all main sub-chain blocks in a view, which would be sorted by their clocks before processing blocks in the new view. Finally, a total order of blocks is retrieved safely. As for transactions, they follow a general order of blocks. For repeated and conflicted transactions, the validity of them is determined by their corresponding blocks and later ones are deemed invalid.

Besides, users of Balloon chain usually need to determine when to confirm a block and transactions within. With reconstruction of $SubTW$ in Balloon, the confirmation policy is similar to GHOST or GHAST \cite{li2020ghast} on a single sub-chain. For example, we could define if a block $b$ is confirmed on the sub-chain if $SubTB(b)$ or $SubTW(b)$ is significantly bigger (e.g., by a certain percent or some number of blocks) than that of its peers (i.e., blocks with the same parent). For genesis after a view change, its peers are genesis blocks on the same sub-chain in the same view. Based on that, we now define, a block $b$ is confirmed on the Ballloon parallel chain only if all main sub-chain blocks in the same view (including $b$) whose clocks are not higher than $b.clock$ are confirmed on their corresponding sub-chains. Many previous works have provided analysis on block confirmation and its safety implications for GHOST-style consensus protocols \cite{sompolinsky2015secure,kiayias2017trees,li2020ghast} , which is therefore omitted here.
\subsection{Safety and Liveness}
This section gives a general discussion on the safety and liveness of Balloon. In the past few years, many researchers have provided their insights on properties of GHOST or its derivative protocols (e.g., \cite{sompolinsky2015secure,li2020ghast,kiffer2018better}). From a certain perspective, Balloon protocol is more like a GHOST version for Parallel chain protocols (with $SubTB$ and $SubTW$ redefined), except that it could dynamically adapts to the network's computation power. As a result, Balloon inherits safety and liveness of GHOST in the first place. Second, block sampling mechanism is applied during block mining for possible chain adjustment. It has no direct impact on the block ordering and thus does no harm to safety. Actually, it is designed for adaptivity, which means the system could adapt to the overall computation power change satisfyingly (e.g., accurately and quickly, $\S5.2$, $\S5.3$). Finally, views in Balloon may be split during the view change process as a result of network asynchrony. However, views across different nodes will eventually merge again with blocks sent later, just like the main chain convergence in GHOST.
\section{Conclusion}
This paper presents Balloon, a scalable parallel chain protocol with good adaptivity. It could adapt its block throughput and performance to the change of computation power. It costs less resource per block and is more economical. Furthermore, Balloon could accomadate small block mining difficulties accessible to general computation hardware, which would make it more open and fair.
\bibliographystyle{plain}
\bibliography{huang2022adaptivity}

\end{document}